\documentclass[english,aps,manuscript,preprint,showkeys]{revtex4}
\usepackage[T1]{fontenc}
\usepackage[latin9]{inputenc}
\usepackage{amsmath}
\usepackage{amssymb}
\usepackage{esint}
\usepackage{babel}
\usepackage{graphicx}

\begin{document}

\title{Lagrange Discrete Ordinates: a new angular discretization for the three dimensional linear Boltzmann
equation}

\author{Cory D. Ahrens}

\affiliation{Colorado School of Mines\\
Department of Applied Mathematics and Statistics\\
Program in Nuclear Science and Engineering\\
1015 14th Street\\
Golden, CO 80401-1887\\
Phone: 720.938.4503\\
Fax: 303.273.3875\\
\texttt{cahrens@mines.edu}~\\
\texttt{\vspace{0.5in}}~\\
Pages: 24 \\
Tables: 0\\
Figures: 4}
\begin{abstract}
The classical $S_n$ equations of Carlson and Lee have been a mainstay in multi-dimensional radiation transport calculations. In this paper, an alternative to the $S_n$ equations, the "Lagrange Discrete Ordinate" (LDO) equations are derived. These equations are based on an interpolatory framework for functions on the unit sphere in three dimensions. While the LDO equations retain the formal structure of the classical $S_n$ equations, they have a number of important differences. The LDO equations naturally allow the angular flux to be evaluated in directions other than those found in the quadrature set. To calculate the scattering source in the LDO equations, no spherical harmonic moments are needed--only values of the angular flux. Moreover, the LDO scattering source preserves the eigenstructure of the continuous scattering operator. The formal similarity of the LDO equations with the $S_n$ equations should allow easy modification of mature 3D $S_n $ codes such as PARTISN or PENTRAN to solve the LDO equations. Numerical results are shown that demonstrate the spectral convergence (in angle) of the LDO equations for smooth solutions and the ability to mitigate ray effects by increasing the angular resolution of the LDO equations.     
\end{abstract}


\maketitle

\section{Introduction}
The discrete ordinates ($S_n$) method can be traced back to the work of Wick \cite{Wick-1943} and Chandrasekhar \cite{Chand-1944}. Both researchers were lead to the discrete ordinate approximation of the one-dimensional, Cartesian geometry transport equation through use of Gaussian quadratures. Because of their use of Gaussian quadratures, one can prove for one-dimensional problems the well known equivalence between the discrete ordinates method and the $P_n$ method, where the angular flux is written as a series of Legendre polynomials. See, for example, Lewis and Miller \citep{LEW-MIL-1985} for a demonstration of this fact. 

A few years later, Carlson and Lee \cite{CAR-LEE-1961} extended the work of Wick and Chandrasekhar to the two- and three-dimensional transport equation. In doing this, they developed the so-called level symmetric quadratures commonly found in many multi-dimensional $S_n$ codes. A fundamental difference, however, between the one-dimensional case initially studied by Wick and Chandrasekhar and the multi-dimensional transport equation is that Gaussian quadratures for the sphere in three dimensions do not exist \cite{MR02}. This lack of higher dimensional Gaussian quadratures prevents the multi-dimensional $S_n$ equations from being equivalent to the $P_n$ equations and, moreover, is at the heart of why ray effects \citep{LATHROP:1968} exist in multi-dimensional $S_n$ solutions. 

In an attempt to improve the $S_n$ method, many researchers have developed new quadratures with various properties. See, for example, \cite{CAR-ZAM-1999, CARLSON-1971, MOREL-89, JAR-ADA-2011, K-K-W-V-1995, SAN-RAG:2011, RUK-YUF-2001, EN-YA-2007, TH-PO-BE-1995} to name just a few. For almost all of the quadratures mentioned, the number of quadrature points (directions on the unit sphere) does not match the dimension of the subspace of spherical harmonics that the quadrature integrates. In fact, this mismatch lead several researchers to expand the subspace of spherical harmonics by adding functions in a somewhat \emph{ad hoc} manner. Morel termed this type of construction as "hybrid collocation-Galerkin" quadratures \cite{MOREL-89}. See also \cite{SAN-RAG:2011}. By obtaining a match between the number of quadrature points and the dimension of the subspace of spherical harmonics it is possible to define Lagrange interpolation on the sphere. We remark that interpolation on the sphere is somewhat more delicate than, for example, one-dimensional interpolation. In one-dimension any $n+1$ distinct points can be used to interpolate a polynomial of degree $n$, but there are sets of points on the sphere with the correct number of points, $\left(N+1\right)^2$ for degree $N$ spherical harmonics, that yield a singular interpolation matrix. Fortunately, these singular point sets are not "common" but highly ill-conditioned interpolation matrices can be constructed if care is not used. Once an interpolatory framework has been established, standard collocation techniques can be used to derive finite dimensional (in angle) approximations to the transport equation.   

In this paper, we take a different approach by leveraging existing interpolatory quadratures, quadratures that yield an interpolation framework, developed by other researchers. Specifically, we use the "extremal" interpolatory quadratures of Sloane and Womersley \cite{SLOA-WOME-2004}, but the results we obtain are not tied to this specific extremal  quadrature. Other interpolatory quadratures, for example, those developed in \cite{FERN-2005} could be used. Using these existing interpolatory quadratures, we derive a new set of "$S_n$ like" equations that we term the "Lagrange Discrete Ordinates" (LDO) equations. The LDO equations, are, by design, quite similar to the classical $S_n$ equations of Carlson and Lee. Thus, existing mature three-dimensional transport codes like PENTRAN \cite{PENTRAN} and PARTISN \cite{PARTISN} can be modified to solve the LDO equations and take advantage of the high performance computing architectures for which these codes have been designed. The LDO equations, however, have a number of important differences, which we now summarize: 
\begin{itemize}
  \item{The solution of the LDO equations has an interpolatory structure in angle that is analogous to one-dimensional Lagrangian interpolation, but on the unit sphere. This allows the angular flux to naturally be evaluated at directions other than the discrete directions from the quadrature set. Moreover, this also opens the door to using angular biasing schemes for hybrid Monte Carlo calculations by solving the corresponding adjoint problem for the LDO equations and then coupling the results to a Consistent Adjoint Driven Importance Sampling (CADIS) framework \cite{WAG-HAG-1998}.}
  \item{To calculate the scattering source in the LDO equations, there is no need to calculate nor store on a three-dimensional spatial grid the spherical harmonic moments of the angular flux. For problems with strong anisotropies, this will be a dramatic savings in memory, since, for a fixed expansion degree $L$, the number of moments that need to be calculated and stored in the traditional $S_n$ equations is $\left(L + 1\right)^2$. Moreover, because there are no spherical harmonics moments to be calculated, parallel performance should increase by reducing communication overhead. Finally, there is no need to assume a rotationally invariant medium, an assumption that is often violated in, for example, photon transport problems in remote sensing applications \citep{SHU-MYN-88}.}
  \item{The positive-weight quadratures sets on which the LDO equations are based can integrate spherical harmonics ranging from degree $0$ to degree $165$. For a fixed maximum degree of integration $L$, the corresponding number of quadrature points (ordinates) is $\left(L + 1\right)^2$. Thus, the quadrature set that integrates degree $L=165$ spherical harmonics contains $27,556$ directions. These points are roughly equally distributed over the sphere and hence this corresponds to approximately $3,445$ directions per octant. We remark that the Gauss-Chebyshev quadrature sets have positive weights and can be generated for any degree of accuracy, but, unlike the quadratures used here, the Gauss-Chebyshev quadratures tend to cluster points near the north and south poles.}
\end{itemize}
Like the classical $S_n$ equations of Carslon and Lee, the LDO equations still suffer from ray effects. This can be attributed to the hyperbolic nature of the approximation to the streaming operator \cite{MO-WA-LO-PA-2003}. However, with the availability of extremely high degree interpolatory quadratures, ray effects can be mitigated by increasing the number of discrete ordinates.

The remainder of this paper is organized as follows. In Section~\ref{sec:background}, the linear Boltzmann equation along with the classical $S_n$ equations are introduced. In Section~\ref{sec:sphere-approximation} the necessary mathematical tools for developing interpolation on the sphere are reviewed and then used in Section~\ref{sec:derivation} to derive the multi-group LDO equations. Here also various properties of the LDO equations are discussed. Preliminary numerical results are presented in Section~\ref{sec:nresults} and finally conclusions and future directions for research are presented in Section~\ref{sec:conclusions}. We remark that this paper is an extension of the results first published in \citep{AHRENS-2012-PHYSOR} to include a multi-group formalism and the presentation of numerical results. 
%
\section{Background}\label{sec:background}
\subsection{Linear Boltzmann equation}
We consider the steady-state, linear Boltzmann equation, written in
standard neutronics notation, 
\begin{align}
  \label{eq:lbe} 
  \boldsymbol{\Omega}\cdot\nabla\psi\left(\mathbf{r},E,\boldsymbol{\Omega}\right) +
    \Sigma_{t}\left(\mathbf{r},E\right)\psi\left(\mathbf{r},E,\boldsymbol{\Omega}\right)= &
    \int_{0}^{\infty}\int_{\mathbb{S}^{2}}\Sigma_{s}\left(\mathbf{r},E'\rightarrow E, \boldsymbol{\Omega}'\cdot\boldsymbol{\Omega}\right)\psi\left(\mathbf{r},E',\boldsymbol{\Omega}'\right)d\Omega'dE' + \\ \nonumber
    + & \,S\left(\mathbf{r},E,\boldsymbol{\Omega}\right).
\end{align}
Equation~(\ref{eq:lbe}) is posed on a convex spatial domain $D\subset\mathbb{R}^3$, $\mathbf{r}\in D$, with energies $0 < E < \infty$ and streaming direction $\boldsymbol{\Omega}\in\mathbb{S}^2$, the unit sphere in $\mathbb{R}^3$. We assume incident flux boundary conditions
\begin{align}
  \psi\left(\mathbf{r},E,\boldsymbol{\Omega}\right)= & \, \Gamma\left(\mathbf{r},E,\boldsymbol{\Omega}\right),\mathbf{\: r}\in\partial D,\:\boldsymbol{\Omega}\cdot\hat{\mathbf{n}}<0, \label{eq:bcs}  
\end{align}
where $\Gamma$ is the prescribed incident flux. The total and scattering cross sections are, respectively, $\Sigma_t$ and $\Sigma_s$ and the external source is described by $S$. We remark that eigenvalue problems can be treated as well, but we limit the presentation to fixed-source problems. As mentioned above, it is not necessary to assume the differential scattering cross section depends on the cosine of the scattering angle $\boldsymbol{\Omega}'\cdot\boldsymbol{\Omega}$, but for simplicity we present only the derivation for a rotationally invariant medium, a case more commonly found in neutron transport problems.

\subsection{The $S_n$ equations}
To facilitate comparison between the LDO equations and the $S_n$ equations, we briefly sketch a derivation of the one-group $S_n$ equations. We start with the mono-energetic transport equation in a homogeneous medium
\begin{align}
  \label{eq:1-grp-lbe} 
  \boldsymbol{\Omega}\cdot\nabla\psi\left(\mathbf{r},\boldsymbol{\Omega}\right) +
    \Sigma_{t}\psi\left(\mathbf{r},\boldsymbol{\Omega}\right)= &
    \int_{\mathbb{S}^{2}}\Sigma_{s}\left(\boldsymbol{\Omega}'\cdot\boldsymbol{\Omega}\right)
    \psi\left(\mathbf{r},\boldsymbol{\Omega}'\right)d\Omega' +
     S\left(\mathbf{r},\boldsymbol{\Omega}\right)
\end{align}
and write the differential scattering cross section in terms of Legendre polynomials,
\begin{equation}
 \Sigma_s\left(\boldsymbol{\Omega}\cdot\boldsymbol{\Omega}'\right) = \sum_{n=0}^{\infty}\frac{2n+1}{4\pi}
 \sigma_s^n P_n\left(\boldsymbol{\Omega}\cdot\boldsymbol{\Omega}'\right), \label{eq:sigma_s_ser-1}
\end{equation}
where $\sigma_s^n$ are the expansion coefficients. The addition theorem for spherical harmonics \cite{LEW-MIL-1985} is now used to convert this expansion to one in terms of spherical harmonics, that is, Eq.~(\ref{eq:sigma_s_ser-1}) is equivalent to
\begin{equation}
 \Sigma_s\left(\boldsymbol{\Omega}\cdot\boldsymbol{\Omega}'\right) = \sum_{n=0}^{\infty}
 \sigma_s^n \sum_{|m|\le n} Y_n^m\left(\boldsymbol{\Omega}\right) \bar{Y}_n^m\left(\boldsymbol{\Omega}'\right), \label{eq:sigma_s_ser-ynm}
\end{equation}
where the over-bar denotes complex conjugation. Substituting this expansion into the scattering integral in Eq.~(\ref{eq:1-grp-lbe}), we obtain the exact equation
\begin{align}
  \label{eq:1-grp-lbe-ynm-src} 
  \boldsymbol{\Omega}\cdot\nabla\psi\left(\mathbf{r},\boldsymbol{\Omega}\right) +
    \Sigma_{t}\psi\left(\mathbf{r},\boldsymbol{\Omega}\right)= &
    \sum_{n=0}^{\infty}\sigma_s^n\sum_{|m|\le n} \phi_n^m\left(\mathbf{r}\right) Y_n^m\left(\boldsymbol{\Omega}\right) +
     \,S\left(\mathbf{r},\boldsymbol{\Omega}\right),
\end{align}
where
\begin{equation}
  \phi_n^m\left(\mathbf{r}\right) \equiv \int_{\mathbb{S}^2} \bar{Y}_n^m\left(\boldsymbol{\Omega}\right) \psi\left(\mathbf{r},\boldsymbol{\Omega}\right) \label{eq:ynm-moments}
\end{equation}
are the spherical harmonic moments of the angular flux. In practice, the scattering kernel is truncated at some finite degree, say $N$. To numerically evaluate the moments in Eq.~(\ref{eq:ynm-moments}), we now use a quadrature, that is, we take discrete points on the sphere $\left\{\boldsymbol{\Omega}_i\right\}_{i=1}^{M}$ and associated weights such that 
\begin{equation}
  \int_{\mathbb{S}^2} \bar{Y}_n^m\left(\boldsymbol{\Omega}\right) \psi\left(\mathbf{r},\boldsymbol{\Omega}\right) \approx \sum_{i=1}^M w_i \bar{Y}_n^m\left(\boldsymbol{\Omega}_i\right)\psi\left(\mathbf{r},\boldsymbol{\Omega}_i\right). \label{eq:ynm-moments-quad}
\end{equation}
To find the values $\psi\left(\mathbf{r},\boldsymbol{\Omega}_i\right)$, Eq.~(\ref{eq:1-grp-lbe-ynm-src}) is evaluated at the quadrature points $\boldsymbol{\Omega}_i$, $i=1,2,\dots,M$  
\begin{align}
  \label{eq:1-grp-lbe-ynm-src-sn} 
  \boldsymbol{\Omega}_i\cdot\nabla\psi_i\left(\mathbf{r}\right) +
    \Sigma_{t}\psi_i\left(\mathbf{r}\right) = &
    \sum_{n=0}^{N}\sigma_s^n\sum_{|m|\le n} \phi_n^m\left(\mathbf{r}\right) Y_n^m\left(\boldsymbol{\Omega}_i\right) +
     \,S\left(\mathbf{r},\boldsymbol{\Omega}_i\right), \,\,i=1,2,\dots,M,
\end{align}
where $\psi_i\left(\mathbf{r}\right) \approx \psi\left(\mathbf{r},\boldsymbol{\Omega}_i\right)$. Equations (\ref{eq:ynm-moments-quad}) and (\ref{eq:1-grp-lbe-ynm-src-sn}) are the $S_n$ equations of Carlson and Lee.  It has been proven that under mild assumptions on the cross sections, that the solution of the $S_n$ equations indeed converges to the solution of the continuous transport equations, Eq.~(\ref{eq:1-grp-lbe}). See, for example, \cite{ANS-GIB-1974} and \cite{MADSEN-1971}. Key to this convergence is the availability of positive weight quadratures on the sphere that can integrate arbitrarily high degree spherical harmonics. An appealing feature of Eqs.(\ref{eq:1-grp-lbe-ynm-src-sn}) is that the "steaming plus collision" operator is diagonal, that is, the left hand side depends only on the single index $i$. This has allowed the development of very efficient methods to solve the $S_n$ equations. We remark that in this derivation there is no functional form assumed for the angular flux.

Before deriving the LDO equations, required material from approximation theory for functions defined on the unit sphere in $\mathbb{R}^3$ will be summarized.

\subsection{Approximation on the sphere\label{sec:sphere-approximation}}

Here we collect relevant facts about approximation theory and spherical
harmonics. Much of the material can be found in \citep{MR02}. We start with the space of square integrable functions on the unit sphere in $\mathbb{R}^{3}$, denoted by $L^{2}\left(\mathbb{S}^{2}\right)$
with $\mathbb{S}^{2}$ the unit sphere in $\mathbb{R}^{3}$. An orthonormal basis for $L^{2}\left(\mathbb{S}^{2}\right)$ is given by the spherical harmonics 
\begin{equation}
  Y_{l}^{m}\left(\theta,\phi\right)=\left(-1\right)^{l}\sqrt{\frac{\left(2l+1\right)\left(l-m\right)!}{4\pi\left(l+m\right)!}}P_{l}^{m}\left(\cos\theta\right)e^{im\phi},\;|m|\le l,\;0\le l,\label{eq:ylm}
\end{equation}
where $P_{l}^{m}$ is the Associated Legendre Function, $\theta$
is the polar angle and $\phi$ is the azimuthal angle. In the sequel,
we will write $Y_{l}^{m}\left(\theta,\phi\right)=Y_{l}^{m}\left(\boldsymbol{\Omega}\right)$,
where the unit vector $\boldsymbol{\Omega}=\left(\sin\theta\cos\phi,\sin\theta\sin\phi,\cos\theta\right)^{T}\in\mathbb{S}^{2}$.
For a given positive integer $L>0$, we define the rotationally invariant
subspace of spherical harmonics $\mathcal{H}_{L}$ as
\begin{equation}
  \mathcal{H}_{L}=\mathrm{span}\left\{ Y_{l}^{m}\,:\,|m|\le l,\,0\le l\le L\right\} ,\label{eq:hl}
\end{equation}
with dimension $d_{L}=\mathrm{dim\left(\mathcal{H}_{L}\right)=\left(L+1\right)^{2}}$.
The space $\mathcal{H}_{L}$ admits a so-called reproducing kernel
given by
\begin{equation}
  K\left(\boldsymbol{\Omega}\cdot\boldsymbol{\Omega}'\right)=\sum_{l=0}^{L}\frac{2l+1}{4\pi}P_{l}\left(\boldsymbol{\Omega}\cdot\boldsymbol{\Omega}'\right),\label{eq:rk}
\end{equation}
where $P_{l}$ is the $l^{\mathrm{th}}$ degree Legendre polynomial
and $K$ satisfies the identity
\begin{equation}
f\left(\boldsymbol{\Omega}\right)=\int_{\mathbb{S}^{2}}K\left(\boldsymbol{\Omega}\cdot\boldsymbol{\Omega}'\right)f\left(\boldsymbol{\Omega}'\right)d\Omega'\label{eq:rk_ident}
\end{equation}
for all $f\in\mathcal{H}_{L}$. We remark that the integral operator
in Eq.~(\ref{eq:rk_ident}) can be thought of as a projection operator
from $L^{2}\left(\mathbb{S}^{2}\right)$ onto $\mathcal{H}_{L}$.

Let $S=\left\{ \boldsymbol{\Omega}_{i}\right\} _{i=1}^{M}$ be a given
set of points (directions) in $\mathbb{S}^{2}$ and assume $M=d_{L}$, that is, the number of directions is equal to the dimension of the subspace $\mathcal{H}_L$. The set $S$ is then said to be a fundamental system of points for $\mathcal{H}_{L}$ if the evaluation functionals 
\begin{equation}
  f\mapsto f\left(\boldsymbol{\Omega}_{i}\right),\: i=1,2,\dots,d_{L},\: f\in\mathcal{H}_{L}\label{eq:eval-functionals}
\end{equation}
are linearly independent. This is equivalent to requiring the interpolation matrix
\begin{equation}
  \mathbf{Y}=\left(\begin{array}{ccccc}
Y_{0}^{0}\left(\boldsymbol{\Omega}_{1}\right) & Y_{0}^{0}\left(\boldsymbol{\Omega}_{2}\right) & Y_{0}^{0}\left(\boldsymbol{\Omega}_{3}\right) & \cdots & Y_{0}^{0}\left(\boldsymbol{\Omega}_{d_{L}}\right)\\
Y_{1}^{-1}\left(\boldsymbol{\Omega}_{1}\right) & Y_{1}^{-1}\left(\boldsymbol{\Omega}_{2}\right)\\
Y_{1}^{0}\left(\boldsymbol{\Omega}_{1}\right) &  & \ddots &  & \vdots\\
\vdots\\
Y_{L}^{L}\left(\boldsymbol{\Omega}_{1}\right) & \dots &  &  & Y_{L}^{L}\left(\boldsymbol{\Omega}_{d_{L}}\right)
  \end{array}\right)\label{eq:interp-matrix}
\end{equation}
to be non-singular. Fundamental systems of points can be constructed in a number of different ways. See, for example, \citep{MR02} or \citep{FERN-2005}. Here we choose to use the so-called ``extremal'' systems of points developed by Sloan and Womersley \citep{SLOA-WOME-2004}. These quadratures are designed by maximizing certain quantities that lead to well conditioned interpolation matrices and have no symmetry conditions  imposed on them. Thus, the extremal point systems used here do not possess symmetries like those of, for example, the commonly used level-symmetric quadratures. However, there are fundamental systems (interpolatory quadratures) that have been constructed where points lie on planes of constant polar angle. See \citep{FERN-2005} for details of their construction.  

With a fundamental system of points $\left\{ \boldsymbol{\Omega}_{i}\right\} _{i=1}^{d_{L}}$, Lagrange functions on the sphere can be defined such that
\begin{equation}\label{eq:lagrange}
  L_{i}\left(\boldsymbol{\Omega}_{j}\right)=\delta_{i,j},\; i,j=1,2,\dots d_{L}.
\end{equation}
Moreover, these functions $\left\{L_i\right\}_{i=1}^{d_L}$ form a basis for $\mathcal{H}_L$. We now define another set of functions $K_{i}\left(\boldsymbol{\Omega}\right)\equiv K\left(\boldsymbol{\Omega}\cdot\boldsymbol{\Omega}_{i}\right)$, $i=1,2,\dots d_L$. Using Eqs.~(\ref{eq:rk_ident}) and (\ref{eq:lagrange}), we find that
the Lagrange functions and the reproducing kernel functions are related by
\begin{equation}
  \int_{\mathbb{S}^{2}}L_{i}\left(\boldsymbol{\Omega}\right)K\left(\boldsymbol{\Omega}_{j}\cdot\boldsymbol{\Omega}\right)d\Omega=\left\langle L_{i},K_{j}\right\rangle =L_{i}\left(\boldsymbol{\Omega}_{j}\right)=\delta_{i,j},\; i,j=1,2,\dots d_{L},\label{eq:bi-ortho}
\end{equation}
where $\left\langle f,g\right\rangle \equiv\int_{\mathbb{S}^{2}}fg\,d\Omega$
is an inner product. Equation ~(\ref{eq:bi-ortho}) indicates that $\left\{ K_{i}\right\} _{i=1}^{d_{L}}$
and $\left\{ L_{i}\right\} _{i=1}^{d_{L}}$ form bi-orthogonal bases
for the subspace $\mathcal{H}_{L}.$ Thus we can write one basis in terms of the other, 
\begin{equation}
  L_{i}\left(\boldsymbol{\Omega}\right)=\sum_{j=1}^{d_{L}}\left\langle L_{i},L_{j}\right\rangle   K_{j}\left(\boldsymbol{\Omega}\right)\label{eq:L-in-t-K}
\end{equation}
and
\begin{equation}
K_{j}\left(\boldsymbol{\Omega}\right)=\sum_{i=1}^{d_{L}}\left\langle K_{j},K_{i}\right\rangle L_{i}\left(\boldsymbol{\Omega}\right).\label{eq:K-in-t-L}
\end{equation}
Define the $d_{L}\times d_{L}$ matrix $\mathbf{L}$ with elements
$\left(\mathbf{L}\right)_{i,j}=\left\langle L_{i},L_{j}\right\rangle$.
Using the addition theorem, the Gram matrix $\mathbf{G}=\mathbf{Y}^{\dagger}\mathbf{Y}$
has the elements $\left(\mathbf{G}\right)_{i,j}=K\left(\boldsymbol{\Omega}_{i}\cdot\boldsymbol{\Omega}_{j}\right)=\left\langle K_{i},K_{j}\right\rangle $.
Eqs.~(\ref{eq:L-in-t-K}) and ~(\ref{eq:K-in-t-L}) then imply that
$\mathbf{L}\mathbf{G}=\mathbf{I}$, where $\mathbf{I}$ is the $d_{L}\times d_{L}$
identity. By construction of the extremal point systems, the matrix
$\mbox{\ensuremath{\mathbf{G}}}$ is well-conditioned and hence the
matrix elements $\left\langle L_{i},L_{j}\right\rangle =\left(\mathbf{G}^{-1}\right)_{i,j}$
can be computed accurately. Moreover, the reproducing kernel can be easily calculated using the three-term recursion relation for Legendre polynomials and hence Eq.~(\ref{eq:L-in-t-K}) gives a convenient way to calculate the Lagrange functions.  

For completeness, we also demonstrate how to change from the Lagrange basis to the spherical harmonic basis.  While the spherical harmonics are naturally indexed with two integers, $l$ and $m$, here it is more convenient to define a single index through the mapping $i = l\left(l + 1\right) + m$, where $0\le l \le L$ and $|m|\le l$. A function $f\in\mathcal{H}_L$, expressed in the spherical harmonic basis can be identified with the vector $\mathbf{f}^{\mathrm{sh}} = \left(f_1, f_2,\dots,f_{d_L}\right)^T$, where $f_i$ is the $i^{th}$ spherical harmonic moment of $f$. Likewise, a function $f\in\mathcal{H}_L$ expressed in the Lagrange basis can be identified with the vector $\mathbf{f}^{\mathrm{lg}} = \left(f_1, f_2,\dots,f_{d_L}\right)^T$, where now $f_i = f\left(\boldsymbol{\Omega}_i\right)$ is the value of $f$ at the point $\boldsymbol{\Omega}_i$. We start with 
\begin{equation}
  f\left(\boldsymbol{\Omega}\right)=\sum_{i=1}^{d_{L}}f_i L_{i}\left(\boldsymbol{\Omega}\right)
  \label{eq:interp}
\end{equation}
and calculate the $i^{th}$ spherical harmonic moment of the right hand side of Eq.~(\ref{eq:interp}) to find, using Eqs.~(\ref{eq:rk_ident}) and (\ref{eq:L-in-t-K}),
\begin{equation}
  \mathbf{f}^{\mathrm{sh}} = \bar{\mathbf{Y}} \mathbf{L}\,\mathbf{f}^{\mathrm{lg}},
  \label{eq:change-of-basis} 
\end{equation}
where the matrix $\mathbf{Y}$ is given in Eq.~(\ref{eq:interp-matrix}), $\left(\mathbf{L}\right)_{i,j}=\left\langle L_{i},L_{j}\right\rangle$ and the over bar indicates complex conjugation. By construction, the two matrices $\bar{\mathbf{Y}}$ and $\mathbf{L}$ are invertible and, more over, are well-conditioned. Thus, Eq.~(\ref{eq:change-of-basis}) provides a formula for changing between the spherical harmonic and Lagrange bases. 

Using the interpolatory framework from above, for any $f\in\mathcal{H}_{L},$
\begin{equation}
  f\left(\boldsymbol{\Omega}\right)=\sum_{i=1}^{d_{L}}f\left(\boldsymbol{\Omega}_{i}\right)L_{i}\left(\boldsymbol{\Omega}\right)
\end{equation}
and hence we can also define a quadrature on $\mathcal{H}_{L}$ by
\begin{equation}
  \int_{\mathbb{S}^{2}}f\left(\boldsymbol{\Omega}\right)d\Omega=\sum_{i=1}^{d_{L}}\int_{\mathbb{S}^{2}}L_{i}\left(\boldsymbol{\Omega}\right)d\Omega\, f\left(\boldsymbol{\Omega}_{i}\right)=\sum_{i=1}^{d_{L}}w_{i}\, f\left(\boldsymbol{\Omega}_{i}\right),
\end{equation}
where $w_{i}=\int_{\mathbb{S}^{2}}L_{i}\left(\boldsymbol{\Omega}\right)d\Omega$
and using Eq. ~(\ref{eq:L-in-t-K}) $w_{i}=\sum_{j=1}^{d_{L}}\left\langle L_{i},L_{j}\right\rangle $.
Sloan and Womersley have found positive weight quadratures for $\mathcal{H}_{L}$,
$L=1,2,3,\dots165$, based on their extremal point systems \citep{SLOA-WOME-2004}. We remark that when $L=165$, there are approximately $3,445$ ordinates per octant. 

%
%
%
%

We briefly summarize the material to this point. The subspace $\mathcal{H}_L$ has (at least) three different basis sets: the spherical harmonics $\left\{Y^m_l,\;|m|\le l,\,0\le l\le L\right\}$, the reproducing kernel functions $\left\{ K_{i}\right\} _{i=1}^{d_{L}}$
and the Lagrange functions $\left\{ L_{i}\right\} _{i=1}^{d_{L}}$. For $\left\{ K_{i}\right\}$ and $\left\{ L_{i}\right\}$ to be a basis, it is critical that the number of directions (quadrature points) is identical to the dimension of the  subspace. Said differently, for a Lagrange or reproducing kernel basis to exist the number of quadrature points must equal $\left(L + 1\right)^2$. This constraint is what precludes the commonly used level-symmetric quadrature sets (and many others) from generating a Lagrange basis. Lastly, there is an explicit change of basis formula for changing between any of the above three bases.

\section{Derivation of new equations}\label{sec:derivation}

Using the interpolatory framework outlined in Section~\ref{sec:sphere-approximation},
we now use a collocation (pseudo-spectral) \cite{GOT-ORS} procedure to derive a discrete (in angle)
approximation to Eq.~(\ref{eq:lbe}). To this end, define
\begin{equation}
  \psi_{L}\left(\mathbf{r},E,\boldsymbol{\Omega}\right)=
  \sum_{i=1}^{d_{L}}\psi_{i}\left(\mathbf{r},E\right)L_{i}\left(\boldsymbol{\Omega}\right),
  \label{eq:psi_L}
\end{equation}
where the coefficients $\psi_{i}\left(\mathbf{r},E\right)$ will be determined through collocation. Substitute Eq.~(\ref{eq:psi_L}) into Eq.~(\ref{eq:lbe}) and define the residual
\begin{align}
  \label{eq:psi_L-teq}
  r_L\left(\mathbf{r},E,\boldsymbol{\Omega}\right) \equiv & \,\boldsymbol{\Omega}\cdot\nabla\psi_{L}\left(\mathbf{r},E,  \boldsymbol{\Omega}\right)+\Sigma_{t}\left(\mathbf{r},E\right)\psi_{L}\left(\mathbf{r},E,\boldsymbol{\Omega}\right) - \\  \nonumber
  - &\int_{0}^{\infty}\int_{\mathbb{S}^{2}}\Sigma_{s}\left(\mathbf{r},E'\rightarrow E,\boldsymbol{\Omega}'\cdot\boldsymbol{\Omega}\right)\psi_{L}\left(\mathbf{r},E',\boldsymbol{\Omega}'\right)d\Omega'dE' - \\ \nonumber
  - & S\left(\mathbf{r},E,\boldsymbol{\Omega}\right).
\end{align}
In general, the residual from approximating the true solution of Eq.~(\ref{eq:lbe}) $\psi\left(\mathbf{r},E,\boldsymbol{\Omega}\right)$ with the finite-dimensional (in angle) approximation $\psi_L\left(\mathbf{r},E,\boldsymbol{\Omega}\right)$ will not be identically zero. We first discretize the energy variable using a standard multi-group approach \cite{LEW-MIL-1985}. Define an energy grid $E_{G}<E_{G-1}<E_{G-2}<\cdots<E_{0}$,
where $G$ is the number of groups. Now define the group constants and the group angular flux as
\begin{subequations}
  \label{eq:group-cnsts}
  \begin{align}
    \Sigma_{t,i}^{g}\left(\mathbf{r}\right) & =\frac{\int_{E_{g}}^{E_{g-1}}\Sigma_{t}\left(\mathbf{r},E\right)\psi_{i}\left(\mathbf{r},E\right)dE}{\int_{E_{g}}^{E_{g-1}}\psi_{i}\left(\mathbf{r},E\right)dE} \label{eq:sig_t_i} \\
    \Sigma_{s,i,i'}^{g'\rightarrow g}\left(\mathbf{r},\boldsymbol{\Omega'}\cdot\boldsymbol{\Omega}\right) & =\frac{\int_{E_{g}}^{E_{g-1}}\int_{E_{g'}}^{E_{g'-1}}\Sigma_{s}\left(\mathbf{r},E'\rightarrow E,\boldsymbol{\Omega'}\cdot\boldsymbol{\Omega}\right)\psi_{i'}\left(\mathbf{r},E'\right)dE'dE}{\int_{E_{g}}^{E_{g-1}}\psi_{i}\left(\mathbf{r},E\right)dE} \label{eq:group-cnst_i} \\ 
    S^{g}\left(\mathbf{r},\boldsymbol{\Omega}\right) & =\int_{E_{g}}^{E_{g-1}}S\left(\mathbf{r},E,\boldsymbol{\Omega}\right)dE \label{eq:src_i} \\
    \psi_{i}^{g}\left(\mathbf{r}\right) & =\int_{E_{g}}^{E_{g-1}}\psi_{i}\left(\mathbf{r},E\right)dE.\label{eq:psi_i}  
  \end{align}
\end{subequations}
Equations~(\ref{eq:group-cnsts}) show that there is an angular dependence to the group cross sections, similar to standard derivations \cite{LEW-MIL-1985}. Instead of the energy dependent angular flux, in practice the energy dependent scalar flux is used to weight the group cross sections. We follow the same approach and take the cross sections to be independent of the index $i$ and hence Eqs.~(\ref{eq:sig_t_i}) and (\ref{eq:group-cnst_i}) reduce to
\begin{subequations}
  \label{eq:group-cnsts-no-i}
  \begin{align}
\Sigma_{t}^{g}\left(\mathbf{r}\right) & =\frac{\int_{E_{g}}^{E_{g-1}}\Sigma_{t}\left(\mathbf{r},E\right)\phi_{L}\left(\mathbf{r},E\right)dE}{\int_{E_{g}}^{E_{g-1}}\phi_{L}\left(\mathbf{r},E\right)dE} \label{eq:sigma_t_no_i} \\
\Sigma_{s,}^{g'\rightarrow g}\left(\mathbf{r},\boldsymbol{\Omega'}\cdot\boldsymbol{\Omega}\right) & =\frac{\int_{E_{g}}^{E_{g-1}}\int_{E_{g'}}^{E_{g'-1}}\Sigma_{s}\left(\mathbf{r},E'\rightarrow E,\boldsymbol{\Omega'}\cdot\boldsymbol{\Omega}\right)\phi_{L}\left(\mathbf{r},E'\right)dE'dE}{\int_{E_{g}}^{E_{g-1}}\phi_{L}\left(\mathbf{r},E\right)dE},
  \end{align}
\end{subequations}
where 
\begin{equation}
  \phi_{L}\left(\mathbf{r},E\right)=\int_{\mathbb{S}^{2}}\psi_{L}\left(\mathbf{r},E, 
   \boldsymbol{\Omega}\right)d\Omega=\sum_{i=1}^{d_{L}}w_{i}\,\psi_{i}\left(\mathbf{r},E\right)\label{eq:scalar-flux}
\end{equation}
is the scalar flux. Integrating Eq.~(\ref{eq:psi_L-teq}) over the
$g^{th}$ group and using Eqs.~(\ref{eq:group-cnsts}) and (\ref{eq:group-cnsts-no-i}) we obtain the $G$ equations
\begin{align}
  \label{eq:mg-te}
  r^g_L\left(\mathbf{r},\boldsymbol{\Omega}\right) = & \boldsymbol{\Omega}\cdot\sum_{i=1}^{d_{L}}\left[\nabla\psi_{i}^{g}\left(\mathbf{r}\right)\right]L_{i}
  \left(\boldsymbol{\Omega}\right) + 
  \Sigma_{t}^{g}\left(\mathbf{r}\right)\sum_{i=1}^{d_{L}}\psi_{i}^{g}\left(\mathbf{r}\right)L_{i}
  \left(\boldsymbol{\Omega}\right) - \nonumber \\ 
  - & \sum_{g'=1}^{G}\sum_{i'=1}^{d_{L}}\left\{ \int_{\mathbb{S}^{2}}\Sigma_{s}^{g'\rightarrow g}\left(\mathbf{r},\boldsymbol{\Omega'}\cdot\boldsymbol{\Omega}\right)L_{i'}\left(\boldsymbol{\Omega'}\right)d\Omega'\right\} \psi_{i'}^{g'}\left(\mathbf{r}\right) - \nonumber \\ 
  - & S^{g}\left(\mathbf{r},\boldsymbol{\Omega}\right),\:g=1,2,\dots G.
\end{align}
Using Eqs.~(\ref{eq:rk_ident}) and (\ref{eq:L-in-t-K}), we now evaluate the scattering integral analytically 
\begin{eqnarray}
\int_{\mathbb{S}^{2}}\Sigma_{s}^{g'\rightarrow g}\left(\mathbf{r},\boldsymbol{\Omega'}\cdot\boldsymbol{\Omega}\right)L_{i'}\left(\boldsymbol{\Omega'}\right)d\Omega' & = & \int_{\mathbb{S}^{2}}\Sigma_{s}^{g'\rightarrow g}\left(\mathbf{r},\boldsymbol{\Omega}'\cdot\boldsymbol{\Omega}\right)\sum_{j=1}^{d_{L}}\left\langle L_{i'},L_{j}\right\rangle K_{j}\left(\boldsymbol{\Omega'}\right)d^{2}\Omega'\nonumber \\
 & = & \sum_{j=1}^{d_{L}}\left\langle L_{i'},L_{j}\right\rangle \int_{\mathbb{S}^{2}}\Sigma_{s}^{g'\rightarrow g}\left(\mathbf{r},\boldsymbol{\Omega}'\cdot\boldsymbol{\Omega}\right)K_{j}\left(\boldsymbol{\Omega'}\right)d^{2}\Omega'\nonumber \\
 & = & \sum_{j=1}^{d_{L}}\left\langle L_{i'},L_{j}\right\rangle \Sigma_{s,L}^{g'\rightarrow g}\left(\mathbf{r},\boldsymbol{\Omega}_{j}\cdot\boldsymbol{\Omega}\right),\label{eq:integral}
\end{eqnarray}
where $\Sigma_{s,L}^{g'\rightarrow g}\left(\mathbf{r},\boldsymbol{\Omega}_{j}\cdot\boldsymbol{\Omega}\right)$
is the scattering cross section restricted to maximum degree $L$. Thus, Eq.~(\ref{eq:mg-te}) becomes 
\begin{align}
  \label{eq:mg-te-2}
  r^g_L\left(\mathbf{r},\boldsymbol{\Omega}\right) = & \,\boldsymbol{\Omega}\cdot\sum_{i=1}^{d_{L}}\left[\nabla\psi_{i}^{g}\left(\mathbf{r}\right)\right]L_{i}
  \left(\boldsymbol{\Omega}\right) + 
  \Sigma_{t}^{g}\left(\mathbf{r}\right)\sum_{i=1}^{d_{L}}\psi_{i}^{g}\left(\mathbf{r}\right)L_{i}
  \left(\boldsymbol{\Omega}\right) - \nonumber \\ 
  - & \sum_{g'=1}^{G}\sum_{j=1}^{d_{L}}\sum_{i'=1}^{d_{L}}\Sigma_{s,L}^{g'\rightarrow g}\left(\mathbf{r}, \boldsymbol{\Omega}_{j}\cdot\boldsymbol{\Omega}\right)\left\langle L_{i'},L_{j}\right\rangle\psi_{i'}^{g'}\left(\mathbf{r}\right) - \nonumber \\ 
  - & S^{g}\left(\mathbf{r},\boldsymbol{\Omega}\right).
\end{align}
The collocation procedure now requires the residual Eq.~(\ref{eq:mg-te-2}) to be zero at the points $\left\{ \boldsymbol{\Omega}_{i}\right\} _{i=1}^{d_{L}}$, which yields the $G\times d_L$ equations
\begin{align}
  \boldsymbol{\Omega}\cdot\nabla\psi_{i}^{g}\left(\mathbf{r}\right) + 
  \Sigma_{t}^{g}\left(\mathbf{r}\right)\psi_{i}^{g}\left(\mathbf{r}\right) = & \sum_{g'=1}^{G}\sum_{j=1}^{d_{L}}\sum_{i'=1}^{d_{L}}\Sigma_{s,L}^{g'\rightarrow g}\left(\mathbf{r},\boldsymbol{\Omega}_{i}\cdot\boldsymbol{\Omega}_{j}\right)\left\langle L_{i'},L_{j}\right\rangle\psi_{i'}^{g'}\left(\mathbf{r}\right) + \nonumber \\
  + & S^{g}\left(\mathbf{r},\boldsymbol{\Omega}_{i}\right),\; i=1,2,\ldots,d_{L},\; g=1,2,\ldots,G.\label{eq:mg-sn-eqs}
\end{align}
Equations ~(\ref{eq:mg-sn-eqs}) are the new multi-group "Lagrange Discrete Ordinate" (LDO) equations.
\emph{Formally}, they are the same as the classical $S_{n}$ equations, but differ in how the scattering source is calculated and in the representation Eq.~(\ref{eq:psi_L}) of the angular flux. This difference in how the scattering source is calculated has important implications, which are discussed below and were mentioned in the Introduction.

The discrete approximation of the boundary condition Eq.~(\ref{eq:bcs}) follows from substituting Eq.~(\ref{eq:psi_L}) for $\psi$ and then evaluating the resulting expression for all incoming discrete ordinates:
\begin{align}
  \psi_{i}^{g}\left(\mathbf{r}\right)= & \, \Gamma^{g}\left(\mathbf{r},\boldsymbol{\Omega}_i\right),\mathbf{\: r}\in\partial D,\:\boldsymbol{\Omega}_i\cdot\hat{\mathbf{n}}<0, \label{eq:bcs-disc}  
\end{align}
 
%

\subsection{Properties of new equations}
We now demonstrate various properties of the LDO equations. To simplify the presentation and focus on the angular discretization, we consider only one energy group, dropping the group index notation, and assume a  homogeneous medium. Defining the $d_L \times d_L$ matrix
\begin{equation}
  \left(\boldsymbol{\Sigma^L_s}\right)_{i,j} \equiv \Sigma_{s,L}\left(\boldsymbol{\Omega}_{i}\cdot\boldsymbol{\Omega}_{j}\right),
\end{equation}
Eqs.~(\ref{eq:mg-sn-eqs}) reduce to 
\begin{equation}
  \boldsymbol{\Omega}\cdot\nabla\psi_{i}\left(\mathbf{r}\right) + 
  \Sigma_{t}\psi_{i}\left(\mathbf{r}\right) =  \left[\boldsymbol{\Sigma^L_s}\mathbf{L}\boldsymbol{\psi}\right]_i + S\left(\mathbf{r},\boldsymbol{\Omega}_{i}\right),\; i=1,2,\ldots,d_{L}, \label{eq:one-group-LDO-eqs}
\end{equation}
where the matrix $\mathbf{L}$ was defined in Section \ref{sec:sphere-approximation} and $\boldsymbol{\psi} = \left(\psi_1, \psi_2,\dots,\psi_{d_L} \right)^T$ is the vector of unknown Lagrange interpolation coefficients. We first recall a property of the continuous scattering operator defined by
\begin{equation}
  \mathcal{K}\psi\left(\boldsymbol{\Omega}\right) \equiv  
  \int_{\mathbb{S}^2} \Sigma_s\left(\boldsymbol{\Omega}\cdot\boldsymbol{\Omega}'\right)\psi
\left(\boldsymbol{\Omega}'\right)d\Omega', \label{eq:sc-operator}
\end{equation}
where we have suppressed the spatial dependence. Again expanding the differential scattering cross section into a Legendre series,
\begin{equation}
 \Sigma_s\left(\boldsymbol{\Omega}\cdot\boldsymbol{\Omega}'\right) = \sum_{l=0}^{\infty}\frac{2l+1}{4\pi}
 \sigma_s^l P_l\left(\boldsymbol{\Omega}\cdot\boldsymbol{\Omega}'\right), \label{eq:sigma_s_ser}
\end{equation}
with $\sigma_s^l = 2\pi\int_{-1}^{1} P_l\left(\mu\right)\Sigma_s\left(\mu\right)d\mu$ and using the addition theorem for spherical harmonics it is a routine calculation to show that the eigenfunctions of the scattering operator are the spherical harmonics $Y_l^m$ with associated eigenvalue $\lambda_l = \sigma_s^l$, that is,
\begin{equation}
  \mathcal{K}Y_l^m\left(\boldsymbol{\Omega}\right) =  
   \int_{\mathbb{S}^2} \Sigma_s\left(\boldsymbol{\Omega}\cdot\boldsymbol{\Omega}'\right)Y_l^m
\left(\boldsymbol{\Omega}'\right)d\Omega' =  \sigma_s^l Y_l^m. \label{eq:sc-eigen}
\end{equation}
Note that for a fixed $l$, there are $2l+1$ linearly independent eigenfunctions $Y_l^m$, $|m|\le l$ that share the same eigenvalue $\lambda_l = \sigma_s^l$. Said differently, the eigenvalue $\lambda_l$ has a geometric multiplicity of $2l+1$.

We now demonstrate that the discrete scattering operator preserves the first $L$ eigenvalues of the continuous scattering operator. To this end, consider an arbitrary function $\psi_L \in \mathcal{H}_L$, written in terms of the Lagrange basis:
\begin{equation}
  \psi_{L}\left(\mathbf{r},\boldsymbol{\Omega}\right)=
  \sum_{i=1}^{d_{L}}\psi_{i}\left(\mathbf{r}\right)L_{i}\left(\boldsymbol{\Omega}\right).
  \label{eq:psi_L_ev}
\end{equation}
We now substitute this representation into the scattering integral and perform calculations similar to those that lead to Eq.~(\ref{eq:integral}). This leads to, again suppressing the spatial notation, 
\begin{equation}
  \tilde{\mathcal{K}}\psi_L\left(\boldsymbol{\Omega}\right) = \sum_{i=1}^{d_L}\psi_i\sum_{j=1}^{d_L}\left\langle L_{i},L_{j}\right\rangle\Sigma_s^L\left(\boldsymbol{\Omega}\cdot \boldsymbol{\Omega}_j\right),
 \label{eq:sc-disc}
\end{equation}
which is just the discrete scattering operator before the collocation procedure has been used. We now use the fact that any $Y_l^m \in \mathcal{H}_L$ can be written as 
\begin{equation}
  Y_l^m\left(\boldsymbol{\Omega}\right)=
  \sum_{i=1}^{d_{L}}Y_l^m\left(\mathbf{r}\right)L_{i}\left(\boldsymbol{\Omega}\right)
  \label{eq:y-lm-L}
\end{equation}
and again use the addition theorem to write the truncated scattering kernel in terms of spherical harmonics to obtain
\begin{equation}
  \tilde{\mathcal{K}}Y_l^m\left(\boldsymbol{\Omega}\right) = \sum_{l'=0}^L \sigma_s^l \sum_{|m'|\le l'} \sum_{i=1}^{d_L}\sum_{j=1}^{d_L}Y_l^m\left(\boldsymbol{\Omega}_i\right)\left\langle L_{i},L_{j}\right\rangle \bar{Y}_{l'}^{m'}\left(\boldsymbol{\Omega}_j\right)Y_{l'}^{m'}\left(\boldsymbol{\Omega}\right).
 \label{eq:sc-disc-2}
\end{equation}
Using the relationship $\mathbf{L}\mathbf{G} = \mathbf{I}$ and the invertibility of the matrix $\mathbf{Y}$, one can show that
\begin{equation}
   \sum_{i=1}^{d_L}\sum_{j=1}^{d_L}Y_l^m\left(\boldsymbol{\Omega}_i\right)\left\langle L_{i},L_{j}\right\rangle \bar{Y}_{l'}^{m'}\left(\boldsymbol{\Omega}_j\right) = \delta_{l,l'}\delta_{m,m'}.
 \label{eq:disc-orth}
\end{equation}
Thus, Eq.~(\ref{eq:sc-disc-2}) reduces to
\begin{equation}
  \tilde{\mathcal{K}}Y_l^m\left(\boldsymbol{\Omega}\right) = \sigma_s^l Y_{l}^{m}\left(\boldsymbol{\Omega}\right),
 \label{eq:sc-disc-3}
\end{equation}
that is, the discrete scattering operator preserves the first $L$ eigenvalues and associated eigenfunctions of the continuous scattering operator.

We now demonstrate that the discrete scattering operator correctly captures ``delta function'' scattering. Starting with Eq.~(\ref{eq:sc-disc}) and using the fact that the coefficients in a Legendre series representation of the delta function are all unity, we find
\begin{equation}
  \tilde{\mathcal{K}}\psi_L\left(\boldsymbol{\Omega}\right) = \sum_{i=1}^{d_L}\psi_i\sum_{j=1}^{d_L}\left\langle L_{i},L_{j}\right\rangle\sum_{l=0}^L\frac{2l+1}{4\pi} P_l\left(\boldsymbol{\Omega}\cdot \boldsymbol{\Omega}_j\right).
 \label{eq:sc-disc-delta-2}
\end{equation}
Then using Eqs.~(\ref{eq:rk}) and (\ref{eq:L-in-t-K}) we obtain the result
\begin{equation}
  \tilde{\mathcal{K}}\psi_L\left(\boldsymbol{\Omega}\right) = \psi_L\left(\boldsymbol{\Omega}\right) .
 \label{eq:sc-disc-delta-3}
\end{equation}
Preservation of the first $L$ eigenvalues of the scattering operator and correctly capturing delta function scattering have important consequences for highly anisotropic transport problems. See, for example, \cite{LAR-MOR:2010} for a discussion of this.

We now present results of preliminary numerical simulations based on the one-group LDO equations.
\section{Numerical results}\label{sec:nresults}

Since the LDO equations are \emph{formally} identical to the classical $S_n$ equations, it is possible in principle to modify any one of the various production three-dimensional transport codes, for example PENTRAN \cite{PENTRAN} or PARTISN \cite{PARTISN}, to solve the LDO equations. To demonstrate basic numerical results, however, a three-dimensional code was written specifically to solve the one-group LDO equations, with prescribed incoming boundary data. The standard diamond-difference method without negative flux fix-up was used to discretize the spatial variables and eight separate transport sweeps were done, along with source iteration without acceleration. The main difference in sweeping for the LDO equations as compared to the classical $S_n$ equations using the level-symmetric quadratures (or any other "octant" based quadrature) is that one does not use the ordinates and weights from one octant to generate (via symmetries) the ordinates and weights for the other octants. Instead, one simply reads in the $(L+1)^2$ ordinates and associated weights and then sweeps based on the sign of the direction cosines. Once one full transport sweep has been completed, the scattering source is updated using matrix multiplication: $\boldsymbol{\Sigma^L_s}\mathbf{L}\boldsymbol{\psi}^n$, with $n$ the source iteration index. Note that regardless of the degree of anisotropy in solution $\psi$ or the scattering kernel, no spherical harmonic moments need be calculated at this step; calculation of the scattering source involves only the values of $\psi_i$, the values of the scattering kernel and the matrix $\mathbf{L}$.

\subsection{Spectral Convergence}
Using the method of manufactured solutions, we demonstrate the expected spectral convergence of the solution of Eqs.~(\ref{eq:one-group-LDO-eqs}), when the solution is smooth (infinitely differentiable). Specifically, we consider a spatially constant, Gaussian in angle solution,
\begin{equation}
  \psi_G\left(\boldsymbol{\Omega}\right) = \exp\left(\frac{||\boldsymbol{\Omega} - \mathbf{e}_z||^2}{4\sigma^2}\right),
  \label{eq:msol}
\end{equation}
where $\mathbf{e}_z$ is the unit vector pointing in the direction of the positive z-axis and we take $\sigma = 1/4$. Substituting $\psi_G$ into Eq.~(\ref{eq:1-grp-lbe}), with spatial distances measured in terms of mean free paths, we find the external source that would generate such a solution to be 
\begin{equation}
  S\left(\boldsymbol{\Omega}\right) = \psi_G\left(\boldsymbol{\Omega}\right) - c\int_{\mathbb{S}^2}P\left(\boldsymbol{\Omega}\cdot\boldsymbol{\Omega}'\right)
  \psi_G\left(\boldsymbol{\Omega}'\right)d\Omega',
  \label{eq:mms-source}
\end{equation}
where $c = \Sigma_s/\Sigma_t$ is the scattering ratio and $P\left(\mu\right)$ is the scattering probability distribution. For the scattering probability distribution, we take the Heney-Greenstein model,
\begin{equation}
  P\left(\boldsymbol{\Omega}\cdot\boldsymbol{\Omega}'\right) = \frac{1}{4\pi}\frac{1-g^2}{\left(1 - 2g\boldsymbol{\Omega}\cdot\boldsymbol{\Omega}' + g^2 \right)^{3/2}},
  \label{eq:HG}
\end{equation}
where $-1\le g \le 1$ is the anisotropy factor \citep{HG:1941}. To evaluate the integral in Eq.~(\ref{eq:mms-source}), we use a high-order quadrature from \citep{AHR-BEY-2009} to calculate the spherical harmonic moments of the Gaussian in Eq.~(\ref{eq:msol}). We then expand the scattering probability distribution function using the addition theorem. The number of spherical harmonic moments needed to accurately represent  Eq.~(\ref{eq:msol}) depends on the value of $\sigma$. For $\sigma=1/4$, the degree $l=28$ spherical harmonic moments are $\sim \mathcal{O}\left(10^{-16}\right)$. For a scattering parameter of $g=0.7$, the $l=28$ coefficient of the Legendre expansion of Eq.~(\ref{eq:HG}) is $\sim \mathcal{O}\left(10^{-5}\right)$. Note that in calculating the integral in Eq.~(\ref{eq:mms-source}), the spherical harmonic moments are multiplied by the Legendre coefficients of Eq.~(\ref{eq:HG}).

\begin{figure}
  \includegraphics[width=\linewidth]{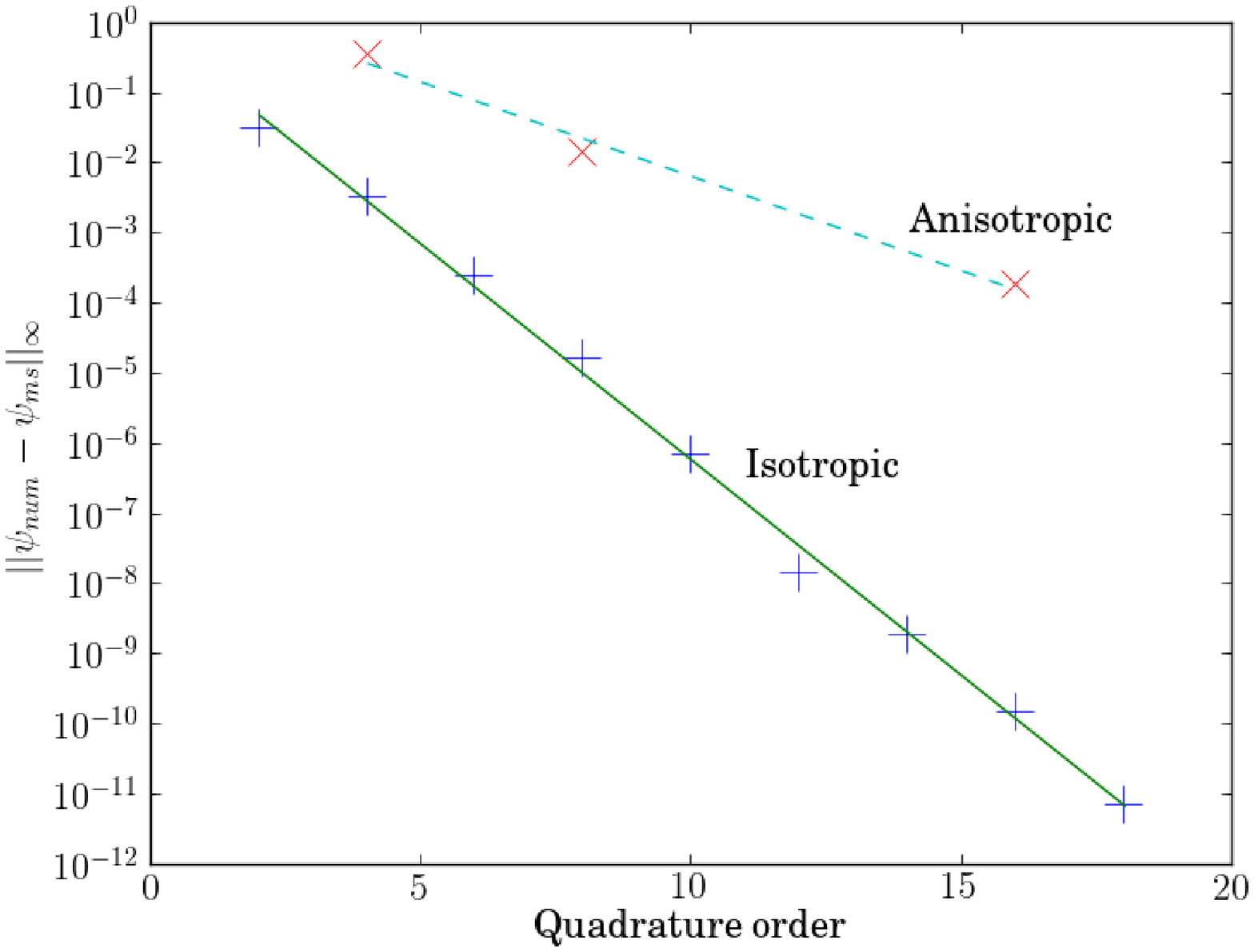}
  \caption{Pointwise, $l_{\infty}$, convergence of numerical solution to $\psi_G$ as a function of the degree of spherical harmonic subspace, $\mathcal{H}_L$ Note that for $\mathcal{H}_L$ there are $\left(L+1\right)^2$ discrete ordinates. The scattering ratio is $c=0.5$. The solid line corresponds to isotropic scattering, while the dashed line corresponds to an $28^{th}$ order expansion of the Heney-Greenstein kernel with $g=0.7$.}
  \label{sconv} 
\end{figure}

Figure ~\ref{sconv} shows the results of using the manufactured source Eq. (\ref{eq:mms-source}) for two cases, one case with isotropic scattering ($g=0$) and one case with forward-peaked scattering ($g=0.7$). The error shown is the maximum point-wise error between the numerical solution of the LDO equations with Eq.(\ref{eq:mms-source}) as the source and the manufactured solution Eq.~(\ref{eq:msol}).
%
%
The linear decrease in error as the number of degrees of freedom increases indicates the expected spectral convergence for an infinitely differentiable solution. For the case of anisotropic scattering, the decrease is not as fast as the isotropic case, since higher degree quadratures (subspaces) must be used to accurately capture the scattering integral. Note that there is no spatial truncation error, since the solution is independent of space. 

\subsection{Point Source}
It is well known that the classical $S_n$ method suffers from ray effects and that they are more pronounced in problems with weak scattering, where particles are only weakly redistributed in angle. Here we demonstrate that, while the LDO equations still have ray effects, they can be mitigated by increasing the number of discrete ordinates. We remark that the points from extremal quadratures sets are nearly evenly distributed over the sphere, which helps to capture spherically symmetric solutions.  

To illustrate this, we simulate a point source located at the center of a spatially uniform box, with the length of each side being five mean free paths. The box is surrounded by vacuum. The uniform spatial mesh has a mesh size of $\Delta x = \Delta y = \Delta z = 0.05$ of a mean free path and the ``point'' source has a volume of $7\Delta x\Delta y\Delta z$. The scattering ratio was taken to be $c=0.25$, and thus there is  strong absorption. 

\begin{figure}[ht]
  \begin{minipage}[b]{0.45\linewidth}
    \centering
    \includegraphics[width=\textwidth]{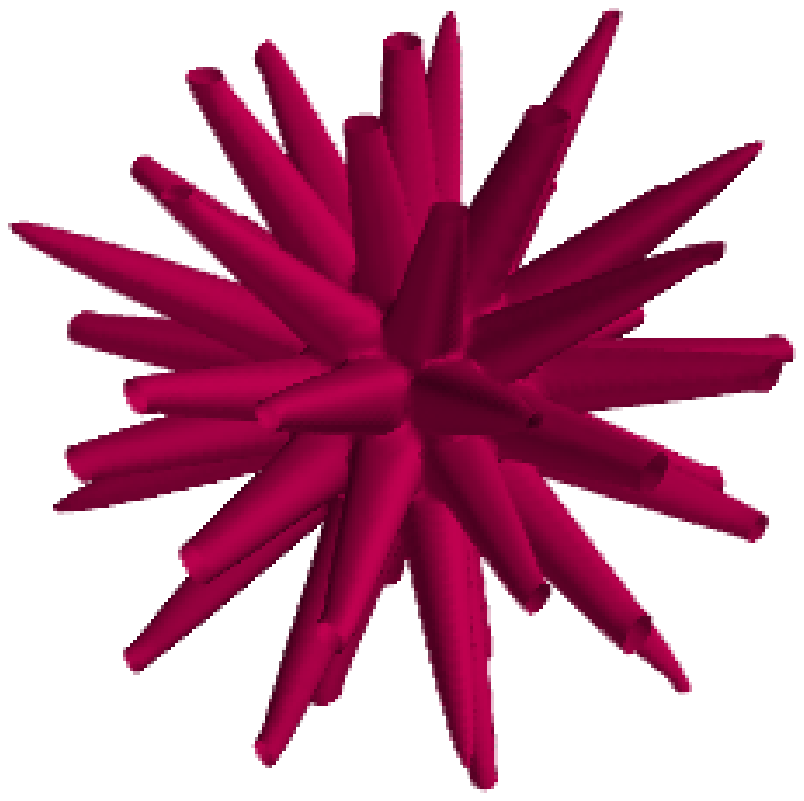}
    \caption{Iso-surface plot of scalar flux from a point-source from solving the LDO equations using a subspace $\mathcal{H}_{6}$ with $49$ points. The scattering ratio $c=0.25$.}
    \label{fig:ldo-l6}
  \end{minipage}
  \hspace{0.1cm}
  \begin{minipage}[b]{0.45\linewidth}
    \centering
    \includegraphics[width=\textwidth]{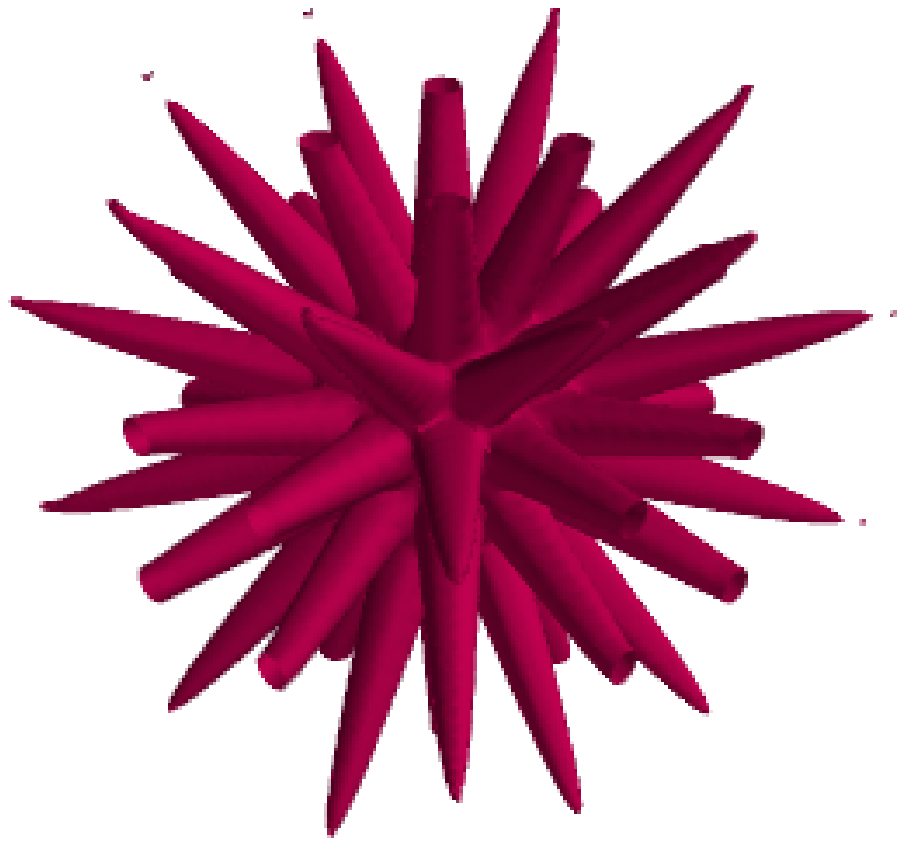}
    \caption{Iso-surface plot of scalar flux from a point-source from PARTISN using a triangular $P_n-T_n$ with $48$ points. The scattering ratio $c=0.25$.}
    \label{fig:s6}
  \end{minipage}
\end{figure}

In Figs.~\ref{fig:ldo-l6} and \ref{fig:s6}, iso-surface plots of the scalar flux are shown. The left figure shows the solution of the LDO equations using the subspace $\mathcal{H}_6$. There are only $49$ discrete ordinates. The right figure shows the solution from PARTISN using a triangular $P_n-T_n$ quadrature, which has $48$ discrete ordinates. While the true solution should be nearly rotationally symmetric, the ray effects are quite pronounced for both solutions.  

\begin{figure}[ht]
  \begin{minipage}[b]{0.45\linewidth}
    \centering
    \includegraphics[width=\textwidth]{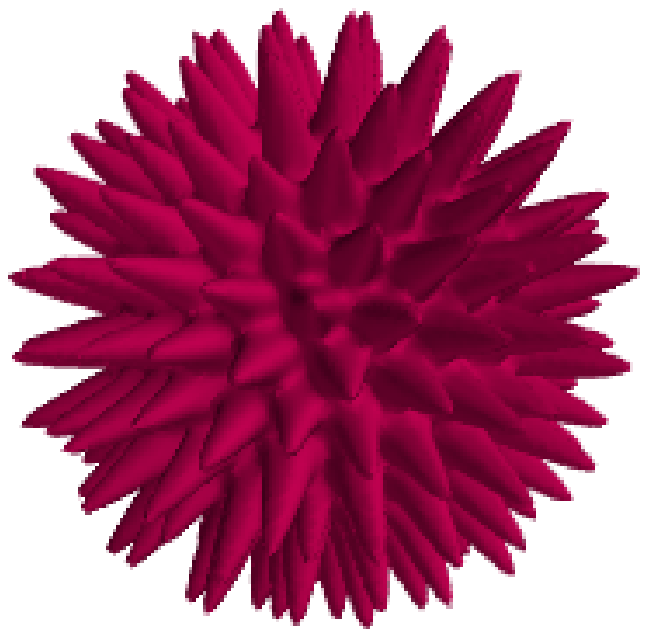}
    \caption{Iso-surface plot of scalar flux from a point-source from solving the LDO equations using a subspace $\mathcal{H}_{12}$ with $169$ points. The scattering ratio $c=0.25$.}
    \label{fig:ldo-l12}
  \end{minipage}
  \hspace{0.1cm}
  \begin{minipage}[b]{0.45\linewidth}
    \centering
    \includegraphics[width=\textwidth]{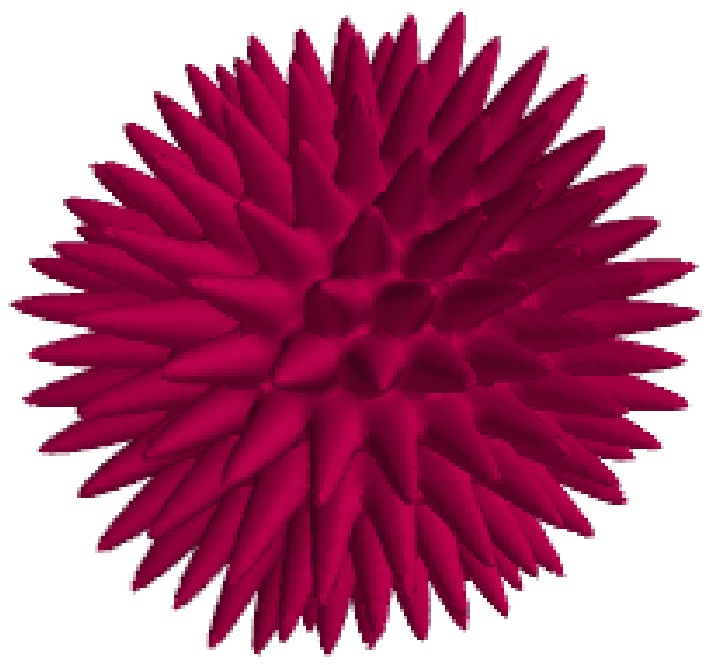}
    \caption{Iso-surface plot of scalar flux from a point-source from PARTISN using a triangular $P_n-T_n$ with $168$ points. The scattering ratio $c=0.25$.}
    \label{fig:s12}
  \end{minipage}
\end{figure}

Figure~\ref{fig:ldo-l12} shows the results of increasing the LDO subspace to $\mathcal{H}_{12}$ with $169$ discrete ordinates, while Fig.~\ref{fig:s12} shows the PARTISN results using a triangular $P_n-T_n$ quadrature, which has $168$ discrete ordinates. While the ray effects are still quite noticeable in both solutions, the solution of the LDO equations appears to be closer to spherically symmetric.   

\begin{figure}[ht]
  \begin{minipage}[b]{0.45\linewidth}
    \centering
    \includegraphics[width=\textwidth]{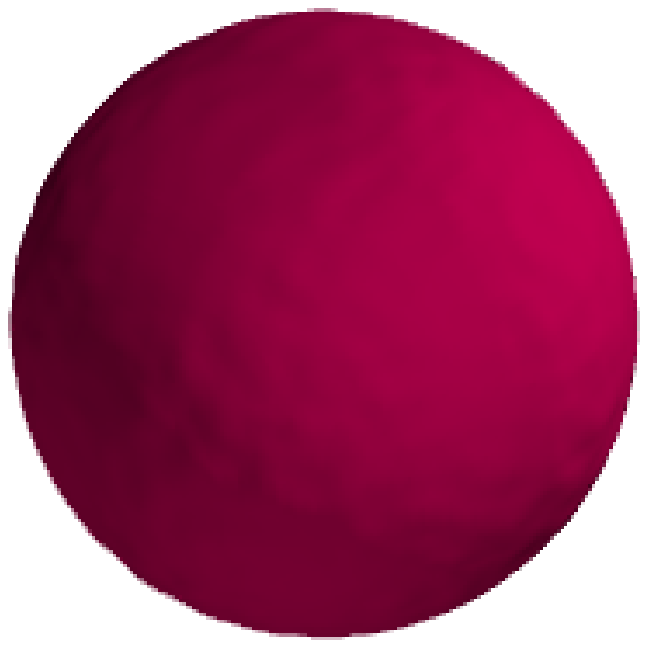}
    \caption{Iso-surface plot of scalar flux from a point-source from solving the LDO equations using a subspace $\mathcal{H}_{28}$ with $841$ points. The scattering ratio $c=0.25$.}
    \label{fig:ldo-l28}
  \end{minipage}
  \hspace{0.1cm}
  \begin{minipage}[b]{0.45\linewidth}
    \centering
    \includegraphics[width=\textwidth]{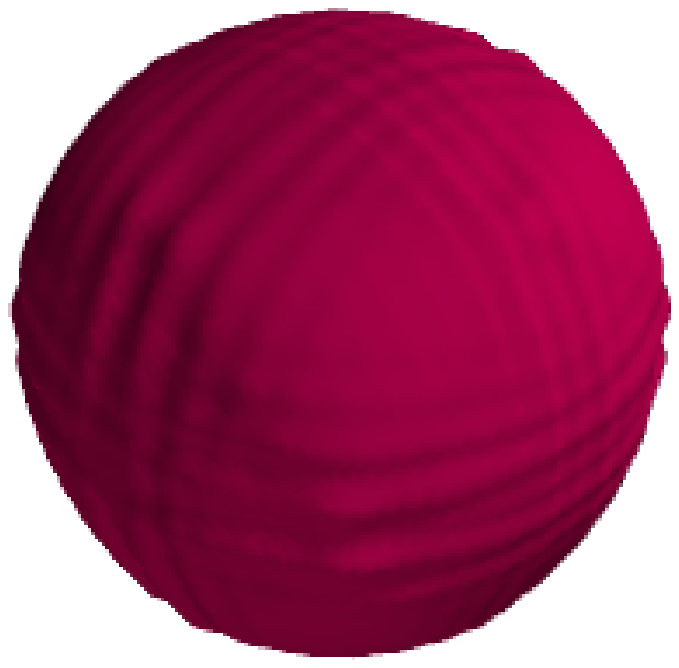}
    \caption{Iso-surface plot of scalar flux from a point-source from PARTISN using a triangular $P_n-T_n$ with $840$ points. The scattering ratio $c=0.25$.}
    \label{fig:s28}
  \end{minipage}
\end{figure}

Figures~\ref{fig:ldo-l28} and \ref{fig:s28} show the results of further angular refinement. The left figure shows the LDO solution corresponding to $\mathcal{H}_{28}$ with $841$ discrete ordinates, while the right figure shows the PARTISN solution with a $840$ point triangular $P_n-T_n$ quadrature. At this angular resolution, the LDO solution appears to be nearly spherically symmetric, while the PARTISN solution shows numerical artefacts of the geometric arrangement of the triangular $P_n-T_n$ quadrature.

\section{Conclusions}\label{sec:conclusions}
By starting with quadrature sets with the correct number of discrete ordinates for an interpolatory framework, we have derived a new set of equations that are {\emph{formally}} the same as the classical $S_n$  equations of Carlson and Lee. We term these new equations the "Lagrange Discrete Ordinate" (LDO) equations. There are a number of notable differences between the LDO and classical $S_n$ equations:
\begin{itemize}
  \item{The LDO scattering source is calculated with only values of the angular flux, with no need to calculate, store and message pass spherical harmonic moments.}
  \item{There is a natural functional representation (in angle) of the numerical solution, which allows one to evaluate the angular flux in directions other than those found in the quadrature sets.}
  \item{The discrete scattering source preserves the first $L$ eigenvalues of the continuous scattering integral, when the subspace $\mathcal{H}_L$ is used to represent the solution $\psi_L$.}
  \item{Currently, extremal interpolatory quadratures exist for $L=1$ up to $L=165$, allowing one to resolve highly anisotropic behavior.}
  \item{Because the LDO equations retain the same structure as the classical $S_n$ equations, much of the existing high performance codes already developed for solving the $S_n$ equations can be easily modified to also solve the LDO equations.}
\end{itemize}
There are a number of directions in which further development of the LDO equations can be taken. First, work is currently under way to modify an existing production-level parallel classical $S_n$ code to also solve the LDO equations. Once this is done, the LDO equations will be benchmarked against community standards. We remark that while in this paper only incoming angular flux boundary conditions were discussed, it is possible to also solve the LDO with reflective boundary conditions, thanks to the interpolatory structure. Numerical results with reflective boundary conditions will be reported in a future publication. In addition, using the change of basis formula Eq.~(\ref{eq:change-of-basis}) one could filter the solution of the LDO equations in manner similar to filtering the $P_n$ equations \cite{MC-HA-LO-2009,AHR-MER-2013}. This should accelerate angular convergence of the LDO equations, allowing the use of lower-order subspaces. Work in this direction is currently underway. 

One drawback with the current formulation of the LDO equations is that to obtain spectral convergence the solution of the transport equation should be smooth in angle. However, it is well known that solutions can actually be quite poorly behaved in angle, thus degrading convergence rates. To address this, one can develop interpolatory frameworks with non-smooth functions \citep{GAN-MHA-2006}. Use of this type of construction for discretizing the three-dimensional transport equation is underway.       

\section{Acknowledgments}
The author would like to thank Professor Glenn Sjoden of Georgia Tech for many interesting discussions on the parallel implementation of $S_n$ equations and for suggestion the name Lagrange Discrete Ordinates.


\clearpage

\clearpage

\clearpage

\clearpage

\end{document}